\documentclass[a4paper]{article}

\usepackage{INTERSPEECH2022}

\title{GLD-Net: Improving Monaural Speech Enhancement by Learning Global and Local Dependency Features with GLD Block}
\name{Xinmeng Xu$^1$, Yang Wang$^2$, Jie Jia$^2$, Binbin Chen$^2$, Jianjun Hao$^{3,*}$\thanks{$^*$ Corresponding Author}}
\address{
  $^1$Electronic \& Elect. Engineering, Trinity College Dublin, Ireland\\
  $^2$vivo AI Lab, China\\
  $^3$School of Foreign Languages, HBUCM, China}
\email{xux3@tcd.ie, $\{$yang.wang.rj, jie.jia, bb.chen$\}$@vivo.com, jianjun$\_$hao2022@163.com}

\begin{document}

\maketitle
\begin{abstract}
For monaural speech enhancement, contextual information is important for accurate speech estimation. However, commonly used convolution neural networks (CNNs) are weak in capturing temporal contexts since they only build blocks that process one local neighborhood at a time. To address this problem, we learn from human auditory perception to introduce a two-stage trainable reasoning mechanism, referred as global-local dependency (GLD) block. GLD blocks capture long-term dependency of time-frequency bins both in global level and local level from the noisy spectrogram to help detecting correlations among speech part, noise part, and whole noisy input. What is more, we conduct a monaural speech enhancement network called GLD-Net, which adopts encoder-decoder architecture and consists of speech object branch, interference branch, and global noisy branch. The extracted speech feature at  global-level and local-level are efficiently reasoned and aggregated in each of the branches. We compare the proposed GLD-Net with existing state-of-art methods on  WSJ0 and DEMAND dataset. The results show that GLD-Net outperforms the state-of-the-art methods in terms of PESQ and STOI.
\end{abstract}

\section{Introduction}
It has been a long-standing topic in speech processing applications to reduce background noise and improve quality and intelligibility of degraded speech. Speech enhancement is frequently used as the preprocessor of the other acoustic tasks, such as speech recognition \cite{ref1}, cochlear implants \cite{ref2}, and hearing aids \cite{ref3}. Monaural speech enhancement provides a versatile and cost-effective approach to the problem by utilizing recordings from a single microphone.

Traditional monaural speech enhancement approaches can include Spectral Subtraction \cite{ref4}, minimum mean-square error (MMSE) estimation \cite{ref5}, Wiener filter (linear MMSE) \cite{ref6}, Kalman filter \cite{ref7}, subspace method \cite{ref8}, etc. However, these kinds of speech enhancement methods generally do not perform well in adverse noise environments.  In recent years, deep neural networks (DNNs) have shown their promising performance on monaural speech enhancement even in highly non-stationary noise environments because of their superior capability in modeling complex non-linearity.

Most of current deep learning architectures for speech enhancement are implemented in time-frequency (T-F) domain by using short-time Fourier transform (STFT), those methods usually treat the spectral magnitude as training target. The phase of noisy speech is utilized along with the enhanced speech magnitude to reconstruct the time-domain signal with inverse short-time Fourier transform (iSTFT) \cite{ref9, ref10, ref11}. Many researchers have investigated the convolution neural network (CNN) or recurrent neural network (RNN). For modeling long-range sequence like speech, CNN requires more convolutional layers to enlarge receptive field. Additionally, although RNN such as long short-term memory (LSTM) or gated recurrent units (GRU) are commonly used in modeling long-term sequence with order information, they cannot perform parallel processing and lead to high computation complexity. Recently, some improvement is achieved by adding temporal convolutional network (TCN) \cite{ref12} blocks or LSTM layers between encoder and decoder for further extracting high-level features and enlarging receptive fields \cite{ref13}. However, the contextual information of observed noisy speech is often ignored, which restricts the denoising performance. 

To tackle the limitations above, we propose a trainable mechanism to infer speech enhancement which contains a two-stage reasoning process according to the human auditory perception. It is noted that human auditory system is able to identify non-speech sounds and human speech through category-selective responses by capturing long-term context-dependency from the scene and related speech regions \cite{ref14}, which was particularly selective for acoustic-phonetic content of speech \cite{ref15}. Considering the concert (i.e. instrument sounds and human talk) as an example, the global-level sound information from the scene (e.g. instrument sounds and mixed speech)  can be first captured as prior knowledge and the speech from the target person is then learned and selected for further processes. Following this perception, we propose a two-stage mechanism to infer speech enhancement where global and local dependency of time-frequency (T-F) bins of speech spectrogram are captured. Aggregating this innovation synthetically, we propose GLD-Net. The contributions of this work are summarized as follows:
\begin{itemize}
    \item A two-stage reasoning mechanism is proposed to learn long-term dependency of T-F bins both in global level and local level from the noisy spectrogram to help detecting interactions between speech part and whole noisy scene. 
    \item An efficient framework for monaural speech enhancement is introduced to efficiently assemble GLD block, namely GLD-Net. The GLD-Net consists of a speech object branch, interference noise branch, and a global noisy branch. In particular, the GLD-Net takes encoder-decoder architecture and the GLD block is inserted in encoder part. In addition, a local-attention is utilized in speech object branch, and interference branch of GLD-Net to detect speech region and filter the noise relevant features, respectively.
    \item To evaluate the proposed GLD-Net, we perform comprehensive ablation studies on public available speech data. Results show that GLD-Net outperforms all 3 evaluated models.
\end{itemize}

\begin{figure}[t]
  \centering
  \includegraphics[width=0.95\linewidth]{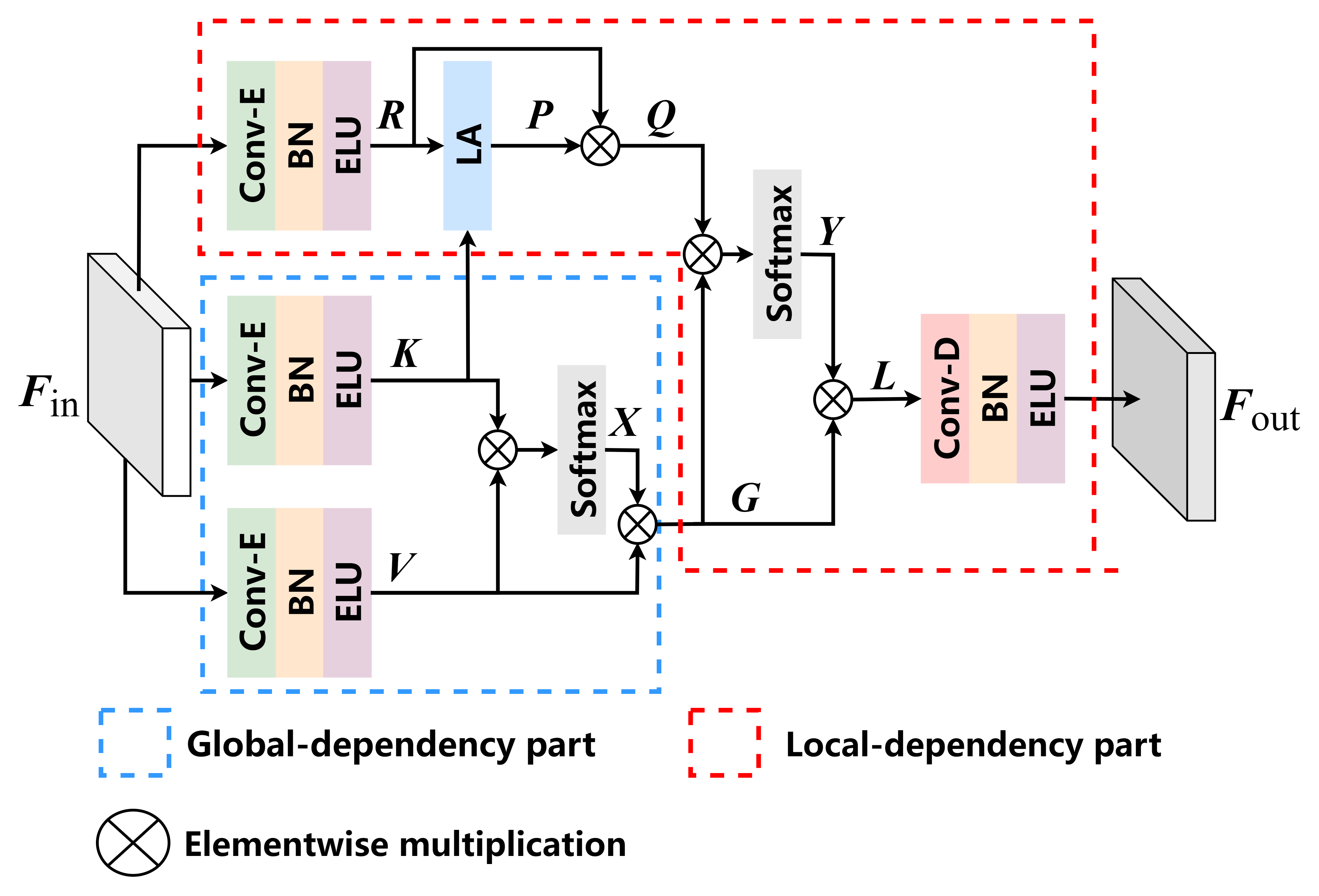}
  \caption{The proposed GLD block in speech branch (one of the three branches from our proposed GLD-Net), where Conv-E, Conv-D, BN, and LA denote convolution encoder, convolution decoder, batch normalization, and local attention, respectively.}
  \label{fig:1}
\end{figure}

\section{Model Architecture}

\subsection{GLD Block}
While listening to a target speech, a person is able to capture its long-range context information from the scene and target speech, rather than just orderly focus on target region of the speech signal like convolution operations do. Thus, we learn from human auditory perception to build a two-stage reasoning mechanism which captures long-range dependency in global-level and local-level to predict target speech. To make the above assumption computationally feasible, we design the GLD block containing two parts to exploit both global and local dependency. Taking speech branch (one of three branches from our proposed GLD-Net) as an example, Figure~\ref{fig:1} shows the detailed procedure of GLD block.

We divide the process of GLD block into two parts, which are global-dependency part and the local-dependency part. While a speech happens throughout the whole noisy observation, we firstly capture global dependency from the whole noisy speech through the global-dependency part. Afterwards, we capture local dependency from speech region through the local-dependency part.

\begin{figure}
  \centering
  \includegraphics[width=0.85\linewidth]{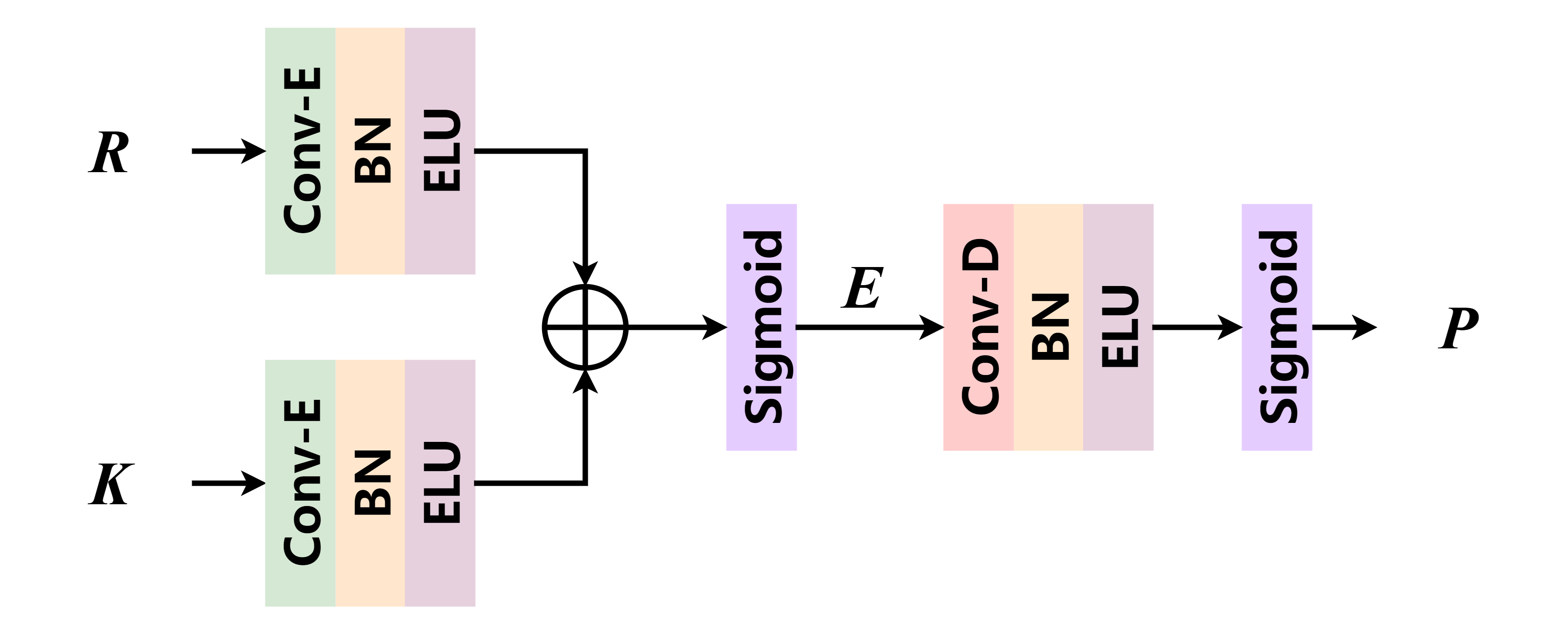}
  \caption{The schematic diagram of local attention.}
  \label{fig:2}
\end{figure}

\subsubsection{Global-dependency part}
In global-dependency part, given an input feature map $F_{\text{in}}$, two 2D convolution layers, each of which is followed by batch normalization \cite{ref16} and exponential linear unit (ELU) \cite{ref17}, are adopt to generate new feature map $\textbf{K}$ and $\textbf{V}$ respectively, where $\{\textbf{K}, \textbf{V}\}\in \mathbb{R}^{T\times F\times C}$, and $T$, $F$, and $C$ denote the number of time steps, frequency bands, and convolution channels. Concretely, a matrix multiplication between $\textbf{K}$, $\textbf{V}^{\top}$ is operated, and to obtain the attention map $\textbf{X}$, a softmax layer is connected as:
\begin{equation}
    x_{ji}=\frac{exp(K_i\times V_j)}{\sum^C_{i=1}exp(K_i\times V_j)},
\end{equation}
in which $x_{ji}$ means the $i^{th}$ channel's influence on the $j^{th}$ channel. The closer feature representation of the two bins signifies a stronger correlation between them.

Finally, the results of matrix multiplication between $\textbf{X}^{\top}$ and $\textbf{V}$ are weighted by a parameter of scale $\alpha$ to acquire the output of global-dependency part $\textbf{G}$:
\begin{equation}
    G_j=\alpha\sum^C_{i=1}(x_{ji}V_j),
\end{equation}
where $\alpha$ is initialized as zero and can be learned gradually.

\begin{figure*}
  \centering
  \includegraphics[width=0.87\linewidth]{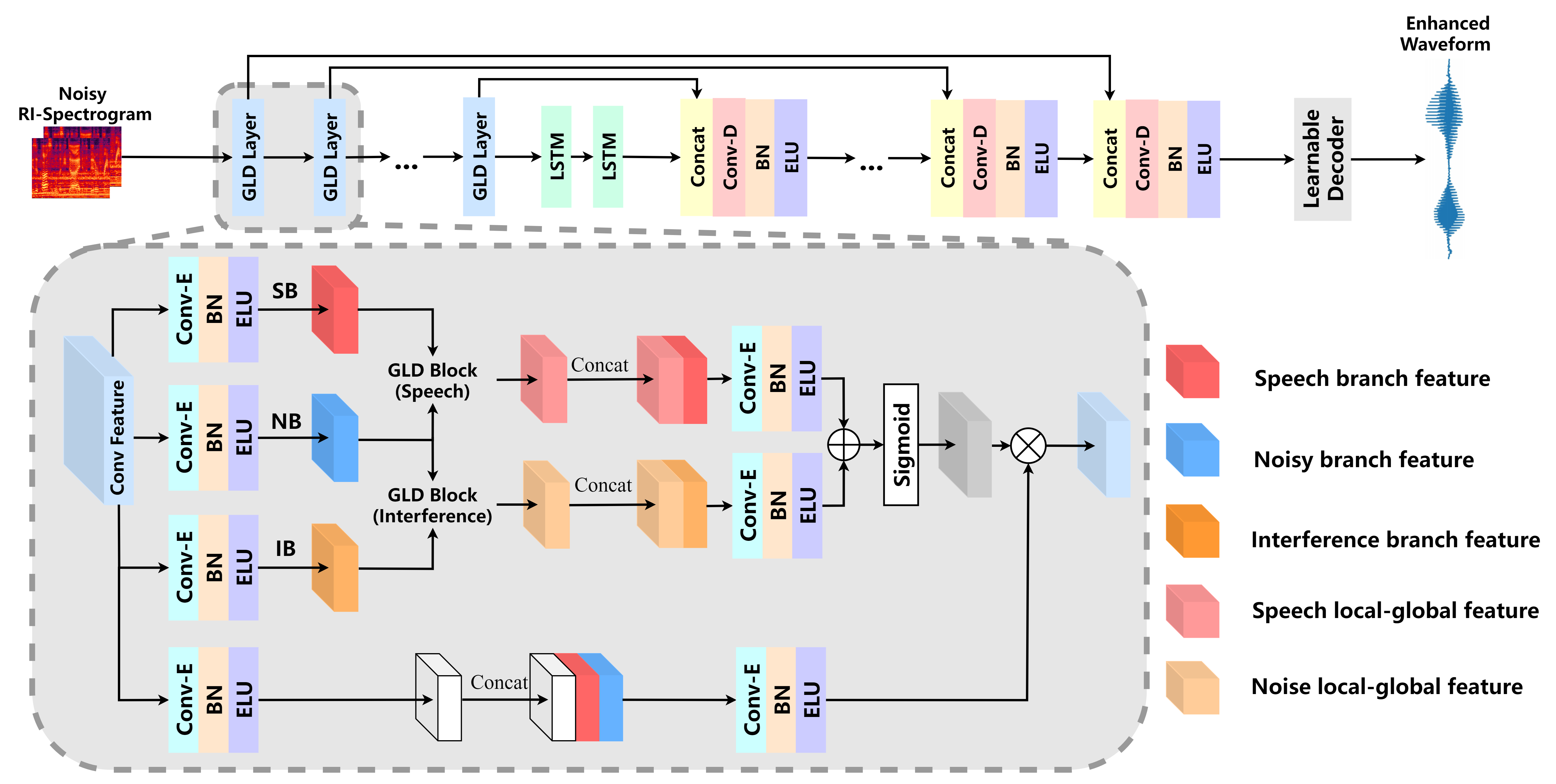}
  \caption{The proposed GLD-Net, where ``SB'', ``NB'', and ``IB'' denote speech branch, noisy branch, and interference branch, respectively.}
  \label{fig:3}
\end{figure*}

\subsubsection{Local-dependency part}
In local-dependency part, the intermediate feature map $\textbf{G}$ is computed from the output of global-dependency part, $\textbf{G}$, and feature map $\textbf{Q}$ which is computed from initial input $F_{\text{in}}$ processing by a 2D convolution block and a local attention (LA).

The LA \cite{ref18} is described in Figure~\ref{fig:2}. $\textbf{E}$ is the $F_{\text{in}}$ processing by the 2D convolution block, and $\textbf{K}$ is the feature map in the global-dependency part. Two additional 2D convolution block, referring as $\textbf{W}_g^{att}$ and $\textbf{W}_x^{att}$, are used to mapping $\textbf{E}$ and $\textbf{K}$ to high-dimensional space feature:
\begin{equation}
    \textbf{R} = \sigma(\textbf{W}_g^{att}\textbf{E}^{\top} + \textbf{W}_x^{att}\textbf{K}^{\top}),
\end{equation}
which is fed into a single 2D deconvolution block, $\textbf{W}_f^{att}$, to give the local attention mask:
\begin{equation}
    \textbf{P} = \sigma(\textbf{W}_f^{att}\textbf{R}^{\top}),
\end{equation}
where $\sigma$ denotes sigmoid function. Then, the feature map $\textbf{Q}$ is generated by
\begin{equation}
    \textbf{Q} = \textbf{R}\times \textbf{P},
    \label{eq}
\end{equation}
The local attention is learning a soft voice activity detection (VAD) and is keeping features focus more on the voice activity region \cite{ref19}. 

Next, a multiplication of metrics is executed between $\textbf{Q}$ and $\textbf{G}$, and a softmax layer is attached subsequently to calculate the attention maps $\textbf{Y}$:
\begin{equation}
    y_{ji}=\frac{exp(Q_i\times G_j)}{\sum^C_{i=1}exp(Q_i\times G_j)},
\end{equation}
where $y_{ji}$ measures the impact of $i^{th}$ bin to the $j^{th}$ bin. Then, a multiplication of metrics is performed between $\textbf{Y}$ and $\textbf{G}$, and the result is:
\begin{equation}
    L_j=\beta\sum^C_{i=1}(y_{ji}G_j),
\end{equation}
where $\beta$ with a zero initial value can be learned to assign larger weight gradually. Finally, the result feature map $\textbf{L}$ is fed into a 2D deconvolution block to form the output of the GLD block, $F_{\text{out}}$. 

The goal of the matrix multiplications and softmax operation in GLD block is to dynamically conduct target-specfic reasoning and capture long-range dependency for a speech sequence. In addition, different T-F bins are considered in the spectrogram from global-dependency part and local-dependency part.

\subsection{GLD-Net}

The proposed GLD-Net uses convolutional encoder-decoder structure based on the standard U-Net \cite{ref20}, shown in Figure~\ref{fig:3}. The encoder contains several GLD layers, which are introduced in Section~\ref{sec221}. The bottleneck consists of two LSTM layers to improve its generalization capabilities \cite{ref21}. The decoder contains several 2D dilated decovolution blocks \cite{ref22}. Skip connections are introduced to compensate for information loss during the encoding process.

In addition, the input of GLD-Net is noisy real-imaginary spectrogram computed by short-time Fourier transform (STFT), denoted by $\textbf{Y}_{r,i}\in \mathbb{R}^{T\times F\times2}$ \cite{ref23}. The output of GLD-Net, $\textbf{S}_{out}$, is processed by a learnable decoder layer to transform the feature back to the predicted time-domain signal, $\textbf{s}_{pred} \in \mathbb{R}^{N\times 1}$. The network structure of the learnable decoder is designed to be exactly the same as the convolutional implementation of inverse STFT. Specifically, the learnable decoder is a 1D transposed convolution layer whose kernel size and stride are the window length and hop size in the STFT \cite{ref24, ref25}.

\subsubsection{GLD layer} \label{sec221}
Each GLD layer of GLD-Net contains three branches, speech branch (SB), noisy branch (NB), and interference branch (IB), as shown in Figure~\ref{fig:3}. The local-level feature (SB feature and IB feature), and the global-level feature (NB feature) are extracted from incoming convolution feature map by using three 2D convolution blocks. Then,  speech local-global feature and noise local-global feature obtained from GLD block are concatenated to SB and IB features, respectively. It is noted that the GLD blocks for SB and IB are different, in which GLD block for SB is as shown in Figure~\ref{fig:1} and GLD block for IB which we change the Eq.(5) of GLD block for SB to $\textbf{Q}=(1-\textbf{P})\times \textbf{E}$. The speech and interference branches takes speech T-F features as inputs to encode long-range dependency.

With the feature vectors, two 2D convolution blocks are employed to fuse the concatenated features, then a matrix addition between the two fused features followed by a sigmoid function are operated to generate local-global dependency gate. Next, several convolution blocks are adopted to extract the feature from original input, obtaining the intermediate feature map. After that SB and NB features are concatenated to intermediate feature, and fed into a 2D convolution block to generate the confidence feature. Our motivation for the concatenation above is that the global context and speech region information are both important for extracting the target speech. Finally, the result of multiplication between confidence feature and local-global dependency gate is the output of a GLD layer, and is passed to the next layer.

\renewcommand{\arraystretch}{1.05}
\begin{table*}[]
\centering
\caption{Evaluation results of proposed model compared with baseline models with unseen speakers. \textbf{BOLD} indicates the best result of each column.}

\begin{tabular}{c|ccccc|ccccl}
\hline
Metric           & \multicolumn{5}{c|}{PESQ}                                                     & \multicolumn{5}{c}{STOI (in \%)}                                                             \\ \hline
Test SNR         & -5            & 0             & 5             & 10            & Avg.          & -5             & 0              & 5              & 10             & \multicolumn{1}{c}{Avg.} \\ \hline
Unprocessed      & 1.37          & 1.88          & 2.12          & 2.49          & 1.97          & 57.06          & 69.33          & 80.59          & 87.63          & 73.40                    \\ \hline
CRN \cite{ref13}             & 2.14          & 2.49          & 2.79          & 3.09          & 2.63          & 78.30          & 88.61          & 91.11          & 94.74          & 88.19                    \\
GRN \cite{ref30}             & 2.20          & 2.58          & 2.84          & 3.12          & 2.69          & 81.45          & 88.09          & 92.43          & 94.96          & 89.23                    \\
T-GSA \cite{ref31}           & 2.33          & 2.70          & 2.96          & 3.24          & 2.81          & 83.17          & 89.54          & 93.47          & 95.76          & 90.49                    \\ \hline
Proposed GLD-Net & \textbf{2.41} & \textbf{2.86} & \textbf{3.03} & \textbf{3.28} & \textbf{2.90} & \textbf{84.06} & \textbf{90.12} & \textbf{94.28} & \textbf{96.19} & \textbf{91.16}           \\ \hline
\end{tabular}
\label{tab:1}
\end{table*}

\renewcommand{\arraystretch}{1.05}
\begin{table*}[]
\centering
\caption{Ablation study of proposed model. \textbf{BOLD} indicates the best result of each column.}
\resizebox{\textwidth}{!}{
\begin{tabular}{c|ccccc|ccccl}
\hline
Metric                                            & \multicolumn{5}{c|}{PESQ}                                                     & \multicolumn{5}{c}{STOI (in \%)}                                                             \\ \hline
Test SNR                                          & -5            & 0             & 5             & 10            & Avg.          & -5             & 0              & 5              & 10             & \multicolumn{1}{c}{Avg.} \\ \hline
Unprocessed                                       & 1.37          & 1.88          & 2.12          & 2.49          & 1.97          & 57.06          & 69.33          & 80.59          & 87.63          & 73.40                    \\ \hline
GLD-Net w/o speech branch                         & 2.23          & 2.59          & 2.86          & 3.11          & 2.70          & 81.36          & 88.41          & 92.21          & 94.86          & 89.21                    \\
GLD-Net w/o interference branch                   & 2.38          & 2.84          & 2.97          & 3.20          & 2.85          & 83.45          & 89.79          & 94.06          & 95.42          & 90.68                    \\
GLD-Net w/o speech and interference branches & 2.17          & 2.48          & 2.77          & 3.04          & 2.62          & 80.32          & 88.24          & 92.05          & 94.76          & 88.84                    \\ \hline
Proposed GLD-Net                                  & \textbf{2.41} & \textbf{2.86} & \textbf{3.03} & \textbf{3.28} & \textbf{2.90} & \textbf{84.06} & \textbf{90.12} & \textbf{94.28} & \textbf{96.19} & \textbf{91.16}           \\ \hline
\end{tabular}}
\label{tab:2}
\end{table*}

\section{Experimental Setup}

In our experiments, we evaluate the models on the WSJ0 SI-84 training set \cite{ref26} including 7138 utterances from 83 speakers (42 males and 41 females). Among these speakers, 6 speakers (3 males and 3 females) are treated as untrained speakers. Hence, we train the models with 77 remaining speakers. To obtain noise-independent models, we select 10 types of noise signals from Demand dataset \cite{ref27}. For training, 4.5 hours clean speech randomly mix with noises at a signal-to-noise ratio (SNR) that is randomly chosen from $\{-5, -3, 0, 3, 5, 7, 10\}$dB. The validation and test set are constructed in the same way, using 1.5 hours of clean speech each, where speakers do not overlap in between the three sets. An additional test sets employing unseen noise types is constructed by using the clean test set speech files and two unseen noise types from the NOISE-92 database \cite{ref28}.  For the evaluation, additional unseen SNR conditions of $\{-5, 0, 5, 10\}$dB are employed.

All data are downsampled to 16 kHz and features are extracted by using frames of length 512 with frame shift of 256, and Hann windowing followed by STFT of size $K=512$ with respective zero-padding. Real and imaginary parts of the complex spectrum are divided into separate feature maps, resulting in $C=2$ input channels. The models are trained with the Adam optimizer \cite{ref29}. The numbers of output channels for the layers in the encoder are changed to 16, 32, 64, 128 and 256 successively, and those for each layer in the decoder to 128, 64, 32, 16 and 1 successively. We set the learning rate to 0.0002. The mean squared error (MSE) is used as the objective function. We train the model with minibatch size of 16.

\section{Results and Analysis}

Evaluation metrics are short-time objective intelligibility measure (STOI) \cite{ref33} and perceptual evaluation of speech quality (PESQ) \cite{ref32}. All metrics are better if higher.

\subsection{Overall Performance}
Table~\ref{tab:1} presents comprehensive evaluations for three baseline models on untrained noises and untrained speakers. The best results in each case are highlighted by boldface. The three selected baseline models are: (1) \textbf{CRN} \cite{ref13}, a CRN speech enhancement model with complex spectral mapping, (2) \textbf{GRN} \cite{ref30}, a gated convolution speech enhancement model with complex spectral mapping, and (3) \textbf{T-GSA} \cite{ref31}, a Gaussian-weighted self-attention based speech enhancement model in frequency domain.

We first compare CRN and GRN, GRN yields significant better STOI and PESQ than CRN. At -5 dB SNR, for example, the GRN improves STOI by $3.15\%$, absolutely and PESQ by $0.06$ over the CRN. It is noted that GRN that takes dilated convolutions, has larger receptive fields than CRN, and thus can leverage long-term contexts. In addition, the T-GSA model substantially outperforms the GRN. Take the -5 SNR case as an example, the T-GSA yields a $1.63\%$ STOI improvement and a $0.13$ PESQ improvement compared with the GRN model. This is observed that transformer neural network promise a stronger long-term dependencies modelling capability than GRN and CRN, which is consistent with the findings in \cite{ref31}.

The proposed GLD-Net consistently outperforms CRN, GRN, and T-GSA in all conditions. Take the $0$ dB SNR case as an example, the proposed GLD-Net improves STOI by $0.58\%$ and PESQ by $0.16$ over T-GSA. These results demonstrate that the proposed GLD-Net which detects the global and local dependency, is more competitive than others in speech enhancement.

\subsection{Ablation Study}
The proposed GLD-Net consists of three branches, SB, NB, and IB, for exploring the the speech local-global dependency and noise local-global dependency. In this ablation study, we evaluate variants of our method when removing SB or IB in this subsection. We set the same parameters for the algorithm with previous section but control the usage of different branches.

As shown in Table~\ref{tab:2}, taking results in $-5$dB SNR case as examples, removing the SB from GLD-Net reduces STOI by $3.74\%$ and PESQ by $0.18$, while removing the IB from GLD-Net reduces STOI by $0.61\%$ and PESQ by $0.04$. That is to say, while adopting just two branches for speech enhancement, SB shows greater strength than IB, which can be interpreted as the speech region contribute that more information than noise region in an observed noisy signal. 

\section{Conclusion}
In this paper, we introduce a two-stage reasoning mechanism for detecting the local and global dependency of speech observation for speech enhancement, namely GLD block. Conforming to the human auditory perception, GLD block not only learns the speech feature from local regions, but also captures both global and local dependency of T-F bins from whole noisy speech RI spectrograms. Without employing any additional inputs, this is a significant try to exploit speech and noise correlation in global-level and local-level in a single trainable mechanism. On this basis, we construct GLD-Net, a three-stream speech enhancement framework consisting of SB, NB, and IB. Through experiments, we show the superiority of the proposed method  over other methods compared in this paper.

\bibliographystyle{IEEEtran}

\bibliography{mybib}


\end{document}